\documentstyle[aps,prl,multicol]{revtex}
\input{epsf}
\begin{document}
\draft
\preprint{\today}
\title{The Kondo screening cloud: 
what can we learn from perturbation theory?}
\author{Victor Barzykin$^1$ and Ian Affleck$^{1,2}$}
\address{Department of Physics$^1$ and Canadian Institute
for Advanced Research$^2$, \\ 
University of British Columbia, Vancouver, BC, V6T 1Z1, Canada}
\maketitle
\begin{abstract}
We analyse the role which the distance scale
$\xi_K = v_F/T_K$ plays in the single-impurity Kondo problem
using renormalization group improved perturbation theory. 
We derive the scaling functions for the local spin susceptibility 
in various limiting cases. 
In particular, we demonstrate exactly that the non-oscillating part 
of it should be short-range, i.e., vanish for distances $r \gg 1/k_F$ 
and show explicitly that the interior of the screening cloud
{\it does not} exhibit weak coupling behavior.\end{abstract}

\pacs{PACS numbers: 
75.20.Hr, 75.30.Mb, 75.40.Cx}
\begin{multicols}{2} 
\narrowtext

  Although the Kondo effect has been very thoroughly studied for
over thirty years\cite{hewson}, comparatively little theoretical
work concentrated on the role of spatial correlations. Part of the
problem is that it is inaccessible to the Bethe ansatz and 
difficult to analyse using Wilson's renormalization group. 
No theoretical and experimental
consensus has emerged on the question  
of the length scale at which screening of the impurity 
takes place.  
On one hand, in the scaling language\cite{wilson,nozieres}, the 
low energy scale $T_K \propto e^{-1/\rho J}$ implies the presence of the
exponentially large length scale $\xi_K = v_F/T_K$; here
$v_F$ is the Fermi velocity, $T_K$ is the Kondo temperature,  
$J$ is the Kondo coupling, and $\rho$ is the density of 
states (per spin). On the other hand, in the Knight shift
experiments of   
Boyce and Slichter\cite{slichter} no such scale was observed.
It  also appears experimentally\cite{hewson} that alloys with 
impurity concentration $n \gg (1/\xi_K)^3$, i.e., where
the inter-impurity distances are much less than $\xi_K$,
display single-impurity behavior. 
In this letter we attempt to clarify these questions using
renormalization group (RG) improved  perturbation theory. 
Related theoretical work includes early perturbative calculations
\cite{kondo}, and RG approaches\cite{chen,gan}.
Recent theoretical work\cite{sorensen,zawadowski} has also addressed
these issues.

In what follows we consider the standard $S_{\rm imp}=1/2$ Kondo model,
\begin{equation}
H =
\sum_{\bf k}\epsilon_k \psi^{\dagger \alpha}_{\bf  k}\psi_{{\bf k}
\alpha} + J{\bf S}_{\rm imp}\cdot \sum_{\bf  k,\bf  k'}\psi^{\dagger
\alpha}_{\bf  k} \frac{\bbox{\sigma}^{\beta}_{\alpha}}{2}\psi_{{\bf  k'}
\beta}.
\label{eq:hkk}
\end{equation}
The quantity measured in the Knight shift experiments is the local
spin susceptibility,
\begin{equation}
\chi (r, T)
\equiv (1/T)<\psi^\dagger ({\bf r}) {\sigma^z \over 2}\psi ({\bf r})
S^z_{\rm tot}>-\chi_0,
\label{kn}
\end{equation}
where $S^z_{\rm tot}=S^z_{\rm imp} +
(1/2) \int d{\bf r}\psi^\dagger ({\bf r}) {\sigma^z}\psi ({\bf r})$ is
the total spin operator of the impurity and conduction electrons. 
The  bulk Pauli contribution,
$\chi_0\approx \rho/2$ has been subtracted.
Recently  a scaling conjecture was made \cite{sorensen}, supported by numerical
results, that  in the scaling limit, $rk_F\gg 1$, $T\ll E_F$,
the spin susceptibility has the following form: 
\begin{equation}
\chi = 
{\chi_{2k_F}\left({rT\over v_F}, {T\over T_K}\right)\over 4\pi^2 r^2v_F}
\cos (2k_Fr)
+{\chi_{un} \left({rT\over v_F}, {T\over T_K}\right)\over 8\pi^2 r^2v_F},
\label{scaling}
\end{equation}
where $\chi_{2k_F}$ and $\chi_{un}$ are universal functions of two
scaling variables\cite{comment}.
This form follows  from 
the relativistic one-dimensional formulation of the
Kondo problem \cite{affleck}. The one-dimensional Hamiltonian in terms of 
the left-moving fields is:
\end{multicols}
\widetext
\begin{equation}
H=v_F\int_{-\infty}^\infty
dr\psi^\dagger_L(r)(id/dr)\psi_L(r) 
 +  v_F\lambda
\psi^\dagger_L(0){\bbox{\sigma} \over 2}\psi_L(0)\cdot
{\bf S}_{\rm imp}.
\end{equation}
\begin{multicols}{2}
\narrowtext
The local spin susceptibility $\chi(r,T)$ in this formalism is a sum
of uniform and $2k_F$ parts.
\end{multicols}
\widetext
\begin{equation}
\chi_{un}(r,T) \equiv
(v_F/T)<[\psi^\dagger_L(r){\sigma^z\over
2}\psi_L(r) \nonumber \\
+\psi^\dagger_L(-r){\sigma^z\over
2}\psi_L(-r)]S^z_{\rm tot}>,
\end{equation}
\begin{multicols}{2} 
\narrowtext
and $\chi_{2k_F}$ is given by the same expression with $r$
replaced by $-r$ in the argument of $\psi_L$.
Here ${\bf S}_{\rm tot}$ is the total spin in the
one-dimensional theory:
\begin{eqnarray}
\label{Stot}
{\bf S}_{\rm tot} &\equiv& {\bf S}_{\rm imp} + {\bf S}_{\rm el} \\
{\bf S}_{\rm el}&=&{1\over 2\pi}\int_{-\infty}^\infty
dr\psi^\dagger_L(r){ \bbox{\sigma} \over 2}\psi_L(r)\nonumber .
\end{eqnarray}  
Renormalizability implies that the functions $\chi_A$ ($A = 2k_F$,
$un$) obey equations of the form:
\begin{equation}
\label{rg}
\left[D{\partial \over \partial D} + \beta(\lambda){\partial
\over \partial \lambda} + \gamma_A(\lambda) \right]
\chi_A(T,\lambda,D, rT/v_F) = 0, \end{equation} 
where $D$ is the ultra-violet cut-off (the bandwidth), $\lambda
\equiv \rho J$ is the dimensionless coupling constant, $\beta (\lambda )$
is the $\beta$-function and $\gamma_A(\lambda )$ is the anomalous
dimension, which is a sum of contributions from the local fermion
bilinears and from $S^z_{\rm tot}$.  In this case, both contributions to
$\gamma_A$ vanish.  $S^z_{\rm tot}$ 
has zero anomalous dimension because it is
a conserved operator.  The fermion bilinear at $r\neq 0$ has vanishing
anomalous dimension because the interactions occur only at the origin;
only ``boundary operators'' receive anomalous dimensions in such
theories \cite{affleck}. The most general solution of this equation is a
general function of the effective coupling constant at scale $T$,
$\lambda_T$ and of the parameter $rT/v_F$.  Swapping the dependence on
$\lambda_T$ for dependence on $T/T_K$ and using the fact that
$\chi_{2k_F}$ is real \cite{sorensen} we obtain Eq. (\ref{scaling}).

\begin{figure}
\centerline{\epsfxsize=2.5in\epsfbox{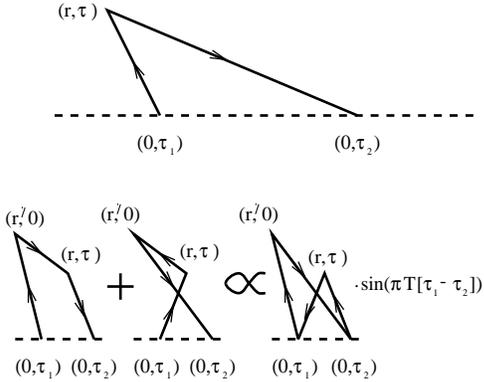}}
\vspace{3mm}
\caption[proof]{Cancellation of the uniform part of the local
spin susceptibility.}
\label{proof}
\end{figure}
Consider first $\chi_{un}$.  
From our lower-order perturbative analysis we have found that 
$\chi_{un}$ is zero. This fact is indeed quite general,
and can be proven to all orders in perturbation theory. 
We will denote as $\chi_{un,\rm{imp}}(r)$ 
the part of the uniform spin susceptibility
for which the impurity piece $S_{\rm imp}$ of $S_{\rm tot}$ 
is responsible, and the other 
part $\chi_{un,\rm{el}}(r)$. [These two contributions are 
defined as zero frequency Fourier transforms.  
Their sum involves a conserved operator,
 $S^z_{\rm imp}$, and so Fourier transforming was equivalent to dividing by
$T$ in Eq. (\ref{kn}).] 
The fact that $\chi_{un,\rm{imp}}(r)$ vanishes can very easily be
seen. In all orders of perturbation theory, one integrates over two free
electron Green's functions, $G$, connecting the points $(r,\tau)$ and $(0,
\tau_i)$ ( see Fig.1a). The result of the $\tau$-integration is zero:
\begin{eqnarray}
I &=& \int_{0}^{\beta} d\tau G(r,\tau-\tau_1)G(-r,\tau_2-\tau) \\
  &=& \int_{0}^{\beta} \frac{d\tau (\pi T)^2}{\sin[\pi T (\tau-\tau_1+ir)]
\sin[\pi T (\tau_2-\tau-ir)]}\,=0 \nonumber,
\end{eqnarray}
because after the change of integration variable, 
$\tau \rightarrow exp(i 2 \pi T\tau)$, one encounters contour integration with
two poles on one side.

The cancellation of $\chi_{un,\rm{el}}(r)$ is somewhat less trivial, since in
addition to the graphs with this type of integration,
there are other graphs. It can be shown that these graphs sum up
to a graph where the integration of the above type is present (see
Fig. 1b) and therefore also vanish. As a result, $\chi_{un}(r,T)$ 
exactly vanishes at finite $r$. This function, however, may include a singular
contribution at $r=0$ (or, more correctly, at $r$ of the order the
ultraviolet cut-off in the one-dimensonal theory, which is essentially
$1/k_F$).

We will now analyse the $2k_F$ part of the spin susceptibility 
$\chi_{2k_F}(r,T)$. From our analysis of 
the perturbation theory up to third order, we
find that the scaling form Eq.(\ref{scaling}) is indeed obeyed. The scaling
function $\chi_{2k_F}$ can be written: 
\end{multicols}
\widetext
\begin{equation}
\chi_{2k_F}\left(x={rT\over v_F},\lambda_T\right) = 
{(\lambda_E+(3\pi/2) \lambda_E^2 x + const \lambda_E^3)(1-\lambda_T) 
\over (4/\pi^2 )\sinh(2\pi x)},
\label{p3}
\end{equation}
where $\lambda_T$ is given by:
\begin{equation} \label{effc}\lambda_T=\lambda +
\lambda^2\ln (D/T)+\lambda^3[\ln^2(D/T)-(1/2)\ln (D/T)+constant].
\label{lambdaeff}\end{equation}   
\begin{multicols}{2} 
\narrowtext
$\lambda_E$ is also given by Eq.(\ref{effc}), with T replaced by another 
effective scale, $E(x) = T/[1-\exp(- 4 \pi x)]e $.  
Eq. (10) is universal up to a rescaling of the
cut-off, $D$ and a change in the $constant$ term. 
 We use this freedom
to redefine $D$ to simplify our expressions.
It is important to
note that the infrared divergences of perturbation theory {\it are not} cut
off at low T by going to small $r$, as was first noticed by
Gan \cite{gan}.  It is also neccessary to have a high $T$ so that
$\lambda_T$ is small.
 In the third order, these
divergences are associated with the graph shown in Fig.2.  Due to the
non-conservation of momentum by the Kondo interaction, the
bubble on the right gives a logarithmic $T$-dependent factor which is
 independent of $r$.
Thus, at low $T$, the interior of the screening cloud does {\it not} exhibit
 weak coupling behavior.
$\chi(r,T)$ decreases exponentially, 
$\propto e^{-2\pi x}$, for $r \gg v_F/T$. 
In the opposite
limit, $r \ll v_F/T$, $\chi_{2k_F}$ can be expressed as a polynomial in
$\lambda_T$ and $\ln x$, or equivalently $\lambda_T$ and $\lambda_r$, the
effective coupling at scale $r$.  
(Note that $\lambda_r \equiv \lambda_E$ for $r \ll v_F/T$.)
To third order, the local spin susceptibility has the
following form:
\begin{equation}
\chi_{2k_F}(\lambda_T,\lambda_r) = {(\lambda_r+c\lambda_r^3)
\over  (r/\pi v_F)}{(1-\lambda_T)\over 4T} \ \ (x \ll 1),  
\label{fac}
\end{equation}
where $c$ is a constant.
The T-dependent factor is,
to the order we work, precisely the total impurity susceptibility,
$\chi_{\rm tt}(T)$.  This is the total susceptibility less the bulk Pauli
term and its value has been determined accurately \cite{hewson}. At low
$T$ it approaches $1/T_K$. $\lambda_r \ll 1$  provided that $r \ll \xi_K$,
even if $T \ll T_K$. 
\begin{figure}
\centerline{\epsfxsize=2.5in\epsfbox{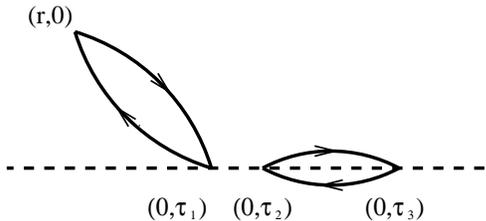}}
\caption{Singular third-order graph for $\chi(r,T)$.}
\label{graph}
\end{figure}
Boyce and Slichter\cite{slichter} have measured the Knight
shift from Cu nuclei near the doped Fe impurities, at distances
up to 5-th nearest neighbor. 
At these very small distances of order of a few lattice spacings,
they have found empirically that the Knight shift obeyed a
factorized form, $\chi(r,T) \approx f(r)/(T+T_K)$,
with rapidly oscillating function $f(r)$ for a
wide range of $T$ extending from well above to well below the Kondo
temperature.  
This is essentially the same
form as Eq. (\ref{fac}) at $T \gg T_K$ and low $r$.
We observe from Eq.(\ref{p3}) that the factorization breaks down 
for $r > v_F/T$. Our perturbative approach isn't valid unless $T \gg T_K$,
so we can't check factorization at low T. 
This question can certainly be addressed for the overscreened large-k
Kondo problem, where it is possible to obtain reliable low-temperature
results from the weak-coupling perturbative expansion \cite{BA}. 
The factorized behavior of the local spin susceptibility was also
obtained in Ref.\cite{cox}.

So far, we have only discussed the region $r \gg 1/k_F$.  It is also
interesting to consider the integral of $\chi (\vec r)$ over all space.
Although the $2k_F$ oscillations tend to cancel at large $r$, there could
be a non-zero contribution from $r \leq 1/k_F$.  
It is useful to consider various
static spin susceptibilities:
\begin{eqnarray}
\label{susc}
\chi_{\rm tt}(T) &\equiv& {<S^z_{\rm tot} S^z_{\rm tot}>\over T}, \ \ 
\chi_{\rm ti}(T) \equiv 
{<S^z_{\rm imp} S^z_{\rm tot}>\over T},\nonumber \\   
\chi_{\rm ii}(T) &\equiv& \int_0^\beta 
<S^z_{\rm imp}(\tau ) S^z_{\rm imp}(0)>  
d\tau .\end{eqnarray}
It is easy to see that the electron spin polarization in the presence
of an impurity is determined by the spatial integral of Eq.(\ref{kn}),
or, equivalently, by $\chi_{\rm tt}(T)-\chi_{\rm ti}(T)$. 
Since $S^z_{\rm tot}$ is conserved, these susceptibilities obey
the RG equation, Eq. (\ref{rg}), with anomalous dimensions determined
by the  anomalous dimension $\gamma_{\rm imp}(\lambda)$ of the operator
$S^z_{\rm imp}$.  For the three different susceptibilities:
 $\gamma_{\rm tt}=0$,
$\gamma_{\rm ti}=\gamma_{\rm imp}$, and $\gamma_{\rm ii} = 2 \gamma_{\rm imp}$. 
The low-order perturbative results for $\beta(\lambda)$\cite{migdal}
and $\gamma_{\rm imp}(\lambda)$\cite{gan} are:
\begin{equation}
\beta(\lambda) = - \lambda^2 + {\lambda^3 \over 2}, \ \ 
\gamma_{\rm imp}(\lambda) = {\lambda^2 \over 2}.
\label{bl}
\end{equation}
We should note that the RG equations are only simple for the
choice of spin correlators Eq.(\ref{susc}). The equations become mixed
if written in terms of impurity, impurity-electron, and electron parts
of the spin susceptibility. 
The general solution of these RG equations has the form:
\end{multicols}
\widetext
\begin{equation}
4 T \chi_j(T, \lambda, \Lambda) = exp\left[\int_{\lambda}^{\lambda_T}
{\gamma_j(\lambda') \over \beta(\lambda')} d \lambda' \right] \Pi_j(\lambda_T)=
\Phi_j(\lambda_T) exp\left[-\int_{0}^{\lambda}{\gamma_j(\lambda') \over 
\beta(\lambda')} d \lambda' \right].
\label{scT}
\end{equation}  
\begin{multicols}{2}
\narrowtext
Here $\Phi_j(\lambda_T)$, $\Pi_j(\lambda_T)$ are some scaling functions.
From our third-order perturbative analysis using Wilson's 
result\cite{wilson} for $\chi_{tt}(T)$ we have obtained that 
the functions $\Phi_j(\lambda_T)$ coincide for all three susceptibilities up
to and including  terms of order $\lambda_T^2$.   When $\lambda$ is 
small, the scale factor in Eq.(\ref{scT}) can be easily calculated from
perturbative expansion of $\beta$ and $\gamma_{\rm imp}$: 
\begin{equation}
\label{scal}
 exp\left[-\int_{0}^{\lambda}{\gamma_{\rm imp}(\lambda') \over
\beta(\lambda')}d \lambda' \right] \simeq 1 + {\lambda \over 2}. 
\end{equation}
Thus, at least at high temperatures, where our perturbative calculation of the
scaling functions is valid, the integrated electronic susceptibility obeys:
\begin{equation}
\int \chi(r,T) d{\bf r} \approx - {\lambda \over 2}  \chi_{\rm{tt}}(T).
\label{conint}\end{equation}  
This result is mostly given by the electron-impurity correlator, while
the electron-electron piece, 
\begin{equation}
\chi_{ee} \equiv \int_0^{\beta} d \tau
\left<S_{el}^z(\tau) S_{el}^z(0) \right> \simeq 
{\lambda^2 \over 4}  \chi_{\rm{tt}}(T),
\end{equation}
is further suppressed by a power of $\lambda$.
In the scaling limit of small bare coupling, $\lambda \rightarrow 0$,  
the total polarization of the conduction electrons 
vanishes, at least at high temperature.  The oscillating funtion $\chi (r)$
integrates to 0, and $\chi_{\rm tt}$ comes entirely from the
impurity-impurity part.  
It should be emphasized, however,
that the result is non-zero at finite bare coupling $\lambda$. (A typical
experimental value of $\lambda$ might be $1/\ln (E_F/T_K)\approx .15$.)  
We conjecture that the equality of the scaling functions,
$\Phi_j(\lambda_T)$ defined in Eq. (\ref{scT})
holds at all T, so that Eq. (\ref{conint}) is true at
all T and small bare coupling. In particular, the integrated electronic
susceptibility then vanishes in the scaling limit of zero bare coupling at
all $T$.  Precisely this result was found at $T=0$ from the Bethe ansatz
\cite{lowenstein}.  
However, this conjecture is {\it not} completely consistent with recent
work of Lesage et al.\cite{Lesage} which extrapolates to the isotropic
Kondo Hamiltonian from an anisotropic model.  While this may well
indicate that our conjecture is wrong, it is also possible that there
is a problem with the extrapolation since we find the
susceptibilities to be very singular in the isotropic limit.  

$\chi_{2k_F}(r,T)$ is a sum of impurity and electron parts,
$\chi_{2k_F,\rm{imp}}$
and $\chi_{2k_F,\rm{el}}$.  While the former obeys the RG equation
Eq.(\ref{rg}) with $\gamma_A=\gamma_{\rm imp}$,
the latter obeys a more complicated
RG equation due to operator  mixing of ${\bf S}_{\rm el}$ and ${\bf S}_{
\rm imp}$. $\chi_{2k_F,\rm{imp}}$ has the same $\lambda$-dependent
factor as in Eq. (\ref{scal}) multiplied by a scaling function:
\begin{equation}
\chi_{2k_F,\rm{imp}}  
\simeq \left(1 + {\lambda \over 2} \right) \chi_{2k_F}^{(1)}(\lambda_T,x),
\end{equation} 
where the scaling function
\begin{equation}
\chi_{2k_F}^{(1)}(\lambda_T,x)=
{(\lambda_E+const \lambda_E^3)(1-\lambda_T)\over
(4/\pi^2 )\sinh(2\pi x)}
\end{equation}
is not the same as in Eq.(\ref{p3}).
Thus, we obtain for all $r$ at weak bare coupling:
\begin{equation} \chi_{2k_F,\rm{el}} \approx 
-{\lambda \over2} \chi_{2k_F}^{(1)}(\lambda_T,x)
+ {(3\pi/2) \lambda_E^2 x(1-\lambda_T) \over
(4/\pi^2 )\sinh(2\pi x)}.
\label{con2k_F}\end{equation}
Hence two different scaling functions are present in the
experimentally measured Knight shift, and their share depends upon
the gyromagnetic ratios for the impurity and conduction
electrons.  Unlike for the total spin susceptibility,
$\chi_{2k_F,el}$ doesn't vanish in the scaling limit of
zero bare coupling. However, it is small compared to $\chi_{2k_F,imp}$
when the bare coupling and $x$ are both small.

We may summarize the response of the weak-coupling Kondo model to a small
magnetic field as follows.  The impurity spin is much more strongly 
affected by a weak Kondo coupling than is the electron gas. 
This is connected with the fact that the free impurity 
susceptibility blows up as $T\to 0$,
whereas the free conduction electron susceptibility does not.  Thus even
a weak Kondo coupling drastically affects $\chi_{\rm tt}$ at low $T$, causing
the diverging Curie susceptibility to level off at $1/T_K$.  
On the other hand the affect on the electron gas can only become 
appreciable in a long distance, infrared limit. The effect of this, 
however, is not very dramatic, because of the factor of $1/r^2$, 
which arises for purely dimensional reasons. 
At short distances, of $O(1/k_F)$,
the excess polarization of the electron gas produced by the Kondo interaction
is small.  This together with the
oscillating nature of the long-distance polarization gives rise to 
a small integrated excess polarization of the electron
gas of  $O(\lambda /T_K)$.

We would like to thank 
J.~Gan, F. Lesage, N.~Prokof'ev, H. Saleur, E.~S.~S\o
rensen,   P.~C.~E.~Stamp, C.~Varma and A. Zawadowski
for useful discussions and comments. 
This research was supported by NSERC of Canada.

\end{multicols}
\end{document}